# Characteristics of Dispersed ZnO-Folic acid Conjugate in Aqueous Medium

Sreetama Dutta, and Bichitra Nandi Ganguly[*]

Applied Nuclear Physics Division, Saha Institute of Nuclear Physics, Kolkata-700064,

*India*

[*] *Corresponding author. Fax: +91-33-2337-4637*

E-mail address:  bichitra.ganguly@saha.ac.in

**Abstract:** The focus of this article is based on the aqueous dispersed state properties of inorganic ZnO nanoparticles (average size ≤ 4 nm), their surface modification and bio-functionalization with folic acid at physiological pH ~ 7.5, suitable for bio-imaging and targeted therapeutic application. While TEM studies of the ZnO nano-crystallites have been performed to estimate their size and morphology in dry state, the band gap properties of the freshly prepared samples, the hydrodynamic size in aqueous solution phase and the wide fluorescence range in visible region have been investigated to establish the fact that the sol is particularly suitable for the bio-medical purpose in the aqueous dispersed state.

**Key words:** ZnO nanoparticle; folic acid; band gap; hydrodynamic size; fluorescence.

## 1. Introduction

Zinc Oxide (ZnO) as semiconductor nano-material has wide range of multipurpose applications due to its unique optical properties, such as wide band gap (~3.37 eV at room temperature) and high exciton binding energy (60 meV) [1]. Therefore, ZnO nanoparticles (NPs) are important for a wide variety of optical methods owing to their size dependent properties that has been nurtured in the recent years [2]. An electronic excitation in semiconductor clusters consisting of a loosely bound electron-hole pairs that is generally delocalized over a NP and the band gap increases as the particle size / the cluster size of the particles diminish (such as in case of the quantum dots, QDs). Interestingly, the trends of semiconductor colloidal systems as luminescence probes for many biological and biomedical applications have become an arena of intense research focus over the past few years [3,4]. ZnO as semiconductor material though environment friendly, but nanometer size particles of the same may offer selective destruction of tumor cells and could be of potential use for drug delivery processes [5] because of there cytotoxic action [6]. Such inorganic NPs need to be properly functionalized for there biological use in aqueous dispersed state, which is presently an active area of research [7] for bio-imaging and drug delivery applications [8].

In many cases, the malignancy of tumors is detected only at advanced stages when chemotherapeutic drugs become increasingly toxic to healthy cells. To improve this condition, both targeted drug delivery [9] and early detection of cancer cells continue to be extensively investigated [10]. Owing to frequent over-expression of folate receptors on the surface of malignant cells, conjugation of cytotoxic agents (could be NPs) to folic acid (FA) have been demonstrated to enhance selective drug delivery to the tumor site [11] and the detection could be achieved through fluorescence imaging or positron emission tomography (PET) techniques. It would be also pertinent to mention here that the size of the nano ZnO particle as QDs could find an easy entry to biological cellular objects in hydrophilic environment. In this regard, mention may be made that the surface chemistry of specially designed QDs (of ZnO NPs) readily lends them to functionalization with targeting proteins/ chemical groups, which may be a key factor to make them benign to normal cells while still retain their cancer targeting properties.

There are however, at least three primary considerations in the application of semi-conductor quantum dots (QDs) for biological applications [12]: (i) increased stability in aqueous environment over a long period of time; (ii) biocompatibility and non-immunogenicity; and (iii) lack of interference with the native properties of nanoparticles.

ZnO is an inexpensive semiconductor luminescent material compared with others [13]. Thus, the objective is to combine semiconductor nanoparticle technology with bioimaging and drug therapy. This approach is expected to have a significant impact on nano-pharmaceutical product development for cancer therapy. In particular, distribution of drug-loaded nano-carriers could be visualized in vivo [11].

In this communication, we primarily emphasize on the physical study of NP [14] as QDs in aqueous phase with nearly monodispersed condition in the nascent stage, its conjugation with a biologically important molecule such as folic acid, the transparent clear dispersion of the bioconjugated ZnO NPs in solution phase along with its optical spectroscopic absorption-emission properties. We envisage such physical properties of ZnO QDs and ZnO-FA will be important for targeted drug delivery processes.

## 2. Materials and methods

### 2.1. Chemical method

Pure ZnO NPs were prepared by the sol-gel technique [14] from zinc acetate (A.R. grade), in alkaline solution phase using 1:1 ammonia solution (Merck, India), at pH ~ 7.5. The Folic acid complexed ZnO (FA-ZnO) samples have been synthesized using Folic acid (M.F: $C_{19}H_{19}N_7O_6$ procured from Sigma Life Science) in the same alkaline solution at room temperature. The requisite amounts of ZnO and FA-ZnO samples have been dissolved in triple distilled water (TDW) for further measurements.

### 2.2. Physical methods of characterization of the ZnO nanoparticles

In order to determine physical properties of the as-prepared material, the particle size has been measured first by transmission electron microscope (TEM) and then the aqueous solvated properties/hydrodynamic size by dynamic light scattering (DLS) method after dilution with TDW as a solvent and as a function of aging. ZnO nanoparticle absorption spectra were measured using ultra-violate (UV/Vis) spectrophotometer. The light absorption properties and the emission properties have also been studied in the UV-Vis range.

The Synthesized ZnO NP samples were examined under TEM in order to study the particle morphology. In the present investigation, the particle morphology has been obtained using Tecnai S-twin, FEI made transmission electron microscope operating at 200 kV, having a resolution of ~ 1 Å. For such analyses, the ZnO sample first has been i) dispersed in TDW and then also in ii) iso-propanol, through a probe sonicator; a drop of the same was placed onto a carbon coated copper grid and dried at room temperature.

X-ray diffraction pattern of the nano crystallites were measured by Seifert XDAL 3000 diffractometer with CuK$_\alpha$ radiation( wavelength of the radiation, k = 1.54 Å) .

Hydrodynamic size of ZnO NPs in aqueous phase was determined by DLS. This measurement of NPs suspensions were obtained using a high performance particle size measuring instruments (Malvern Instruments Ltd., Worcestershire, U.K.), with He- Ne Laser of wavelength 633 nm. Appropriate amounts of nano-ZnO sample were added to TDW medium. All suspensions were vigorously shaken using sonicator to break up visible clumps to yield a clear dispersion.

UV visible spectroscopy of the electronic transitions of molecules was carried out using a UV-VIS-NIR scanning spectrometer (Lamda 750, Perkin Elmer). The optical absorption spectra were measured in the range of 250–800 nm. From the absorption spectrum we measure the band gap of dispersed NPs in the system.

The steady-state fluorescence measurements have been made in a Spex vluoromax-3 spectrofluorimeter. A pair of 1 cm×1 cm path length quartz cuvettes were used for absorption and emission experiments of the dispersed solutions.

## 3. Results and discussion
### 3.1. Morphological Investigation by TEM

The microstructures of pure ZnO and FA-ZnO NPs have been investigated by TEM under high vacuum and under dry conditions. Typical morphology of synthesized pure ZnO NPs has been shown in Fig. 1.a). as prepared in iso-propanol (since ZnO NPs are insoluble) and they are in dry condition. The electron diffraction pattern and fringe structure of the NPs in 1(b). A swollen structure has been observed in case of FA-ZnO sample (shown in Fig. 1(c)). Hexagonal wurtzite structure has been shown in freshly prepared pure ZnO sample1(d) through X-ray diffraction pattern. A histogram has been constructed (shown in Fig. 2) for synthesized pure ZnO and FA-ZnO samples. From the histograms we can have an average estimate the particle size ≤ 4 nm for synthesized pure ZnO and ~ 11 nm for FA-ZnO sample. It also shows that the particle size is nearly monodispersed. Such small size range of NPs are useful for cancer cell selectivity[8].

*3.2. Dynamic light scattering (DLS) measurement*

In solution phase, DLS results for the hydrated ZnO NPs explored in this study are shown in Fig. 3. Dynamic light scattering method is used to measure Brownian motion of colloidal dispersed particles and to relate this to the hydrodynamic size of the particles in the dispersed solution through dynamic fluctuations of scattered light intensity. This scattered light intensity is further mathematically manipulated to relate the hydrodynamic size of the particles. An important feature of Brownian motion measured by DLS is that small particles move faster in comparison to large particles, and the relationship between the size of a particle and its speed due to Brownian motion is defined in the Stokes-Einstein equation [15]. The fundamental size distribution determined by DLS is a scattered light intensity distribution, which can be converted to a volume distribution using Mie theory [16]. It actually measures the hydrated sphere of the molecules around the ZnO or FA-ZnO samples, which attains stability with time. Factors that affect the diffusion speed of the NPs here are (i) ionic strength of the medium, (ii) surface structure, by changing the thickness of electro chemical double layer/ Debye layer ($\kappa^{-1}$) (Fig. 4) [17]. In our method, we had dispersed the ZnO NPs in TDW which could produce an extended double layer of polarized water (H-O-H) molecules around the particle. Also, any change due to surface conjugation of ZnO-NPs, due to large folic acid molecule affects the diffusion speed and will automatically affect the size of the NPs. The freshly prepared samples (both pure ZnO and FA-ZnO samples in TDW) have shown considerable small hydrodynamic size (2R ~ 388 nm for pure ZnO and ~ 518 nm for ZnO/FA, shown in the Table 1). However, the size of the NPs are very much higher than TEM nano particles, as the particles are isolated from the medium immediately after sonication and dried and put to high vacuum for investigation. In solution phase however, this discrepancy is obvious as pure ZnO and FA-ZnO NP samples further tend to get hydrated and agglomerated with time when dispersed in water [18]. It is appropriate to mention that electrostatic (charge) characteristics of ZnO NPs are important feature for bio-applications (it is found to be pH dependent[19]), in pure aqueous phase they have neutral hydrated groups attached thus dispersed ZnO has hydrated cell around it (shown in the Fig. 4). The Debye length(($\kappa^{-1}$), in this context refers to the mean radius of the oriented shell of water dipoles (neutral) around the NP. Since κ is proportional to the square root of ionic strength of the medium, shell structure diminishes under ionic conditions. The solvated interaction of ZnO NPs in aqueous medium leaves its impact as this type of effect is persistent also from Fig.1a., although it has been dried under vacuum.

This electrostatic interaction of NPs is physiologically important, as it drives various biological interactions. In fact, DLS results show this effect through the hydrodynamic size of the dispersed particles. In case of FA-conjugated particles at physiological pH~7-7.5 of the medium, it is this electrostatic effect that drives the (negative ends) dipoles of FA molecules to orient themselves towards the ZnO particles through their surface positive charge. It may be mentioned that cancer cells frequently contain a high concentration of anionic phospholipids on their outer membrane and large membrane potentials[20], thus electrostatic interactions of bare ZnO NPs for such biologically driven sites deserves special mention in this context.

Ultimately, we obtain a clear transparent solution on standing. Hydrodynamic sizes (2R) of both samples have been increased due to aging as is evident from Table 1. No physical change has been observed after one month time and we consider the solution phase to be stable.

*3.3. UV-Vis spectrum analysis*

Tailoring the band structure is a vital assignment for the research of semiconductor material, like ZnO and this property lies at the root of many physico-chemical processes of the cytotoxic properties of NPs. The electron and hole pair generated in these semiconductor NPs mediated processes can participate in many kinds of redox reaction cycles to generate reactive oxygen species (ROS) in cellular environments [5,21]

The band gap or band structure of the material is found to be size dependent owning to the well-known quantum confinement effects. According to Beer-Lambert's law, the absorbance (A) is related by the equation

$$A = \log(I_0/I) = \alpha d \qquad (i)$$

where, $I_0$ and I are the intensity of incident and transmitted light respectively, $\alpha$ is the absorption coefficient (= ε.c; where, ε is the extinction coefficient and c is concentration) and *d* is the path length of absorbing solution (in this case, *d* =1 cm and concentration is uniform).

ZnO is a direct band semiconductor for which $\alpha$ is related to the excitation energy

$$\alpha E = A(E - E_g)^{1/2} \qquad (ii)$$

where, $E$g is a band gap energy. Standard extrapolation of absorption onset (as shown in Fig. 5) [22] to $\alpha E = 0$ (where $E = E$g) has been made for each samples. The band gap value of freshly prepared sample, shown in Fig. 6, has been estimated ~ 3.9 eV, which is quite high compared to standard sample [1,2]. S. Monticone et al. [2] has reported the band gap of freshly prepared sample as 3.75 eV. It has been also shown that the band gap value of ZnO NPs changes due to aging in solution [2]. Besides this, the band gap value has been changed with pH [23]. Seema Rani et al. [23] estimated the band gap value ~ 3.5 eV due to pH ~ 7 for the particle size ~ 7.5 nm. In the present study, we have estimated the band gap ~ 3.9 eV for pH ~ 7.5 and particle size ~ 4 nm (from TEM result). From the Table 2 we observed that the band gap value has been tuned from 3.9 eV to 2.8 eV with increasing Folic acid concentration. The band gap effect is reduced due to donor charge level since ZnO is ensconced by the FA [14]. According to Brus [24] the change in band gap energy for 3d direct band gap material, Eg, is given as:

$$\Delta E_g = \frac{\hbar^2 \pi^2}{2\mu D^2} - \frac{1.8 e^2}{\varepsilon_2 D} \qquad (iii)$$

where D = 2R, μ is reduced mass and $\varepsilon_2$ is dielectric constant.

The quantum confinement is material specific because it depends on dielectric constant of the material. The analysis seems to work particularly well for ZnO particles when the size gets smaller than 5-6 nm. In this case, the NPs average size obtained is ~ 4 nm. Therefore, quantum confinement effect is observed and we find a considerable increase of band gap energy. But the band gap effect decreases due to FA ensconced charge at donor level (Schematic diagram shown in Fig. 7) with the surface modification of the NPs.

*3.4. Analysis of emission spectrum*

Photo activation of ZnO NPs has been effectively utilized in generating greater levels of ROS release which if effectively targeted to cancer cells will lead to selective destruction[5,20]. Recent reports reveal that illumination or photo-excitation of ZnO conjugated ZnO NPs resulted in synergistic cytotoxic action in ovarian cancer cells, where as little cytotoxicity has been observed under dark condition[24]. We found synthesized pure ZnO and FA-ZnO samples dispersed in aqueous medium show a clear emission spectrum in the region 400-500 nm in Fig. 8. Folic acid (concentration) ensconcement can help to overall enhancement of the intensity of florescence emission[25], the emission peak $\lambda_{max}$ ~ 450 nm (slightly red shifted in both the cases than pure ZnO). In case of excitation energy at 365 nm, in FA-ZnO samples the emission is increased by two times, which means ZnO in organic confinement and FA templated structure gives better fluorescence as shown from the Fig. 8(b). This property can be utilized in cytotoxic action to generate ROS cascade in cellular environment.

Finally it needs to be mentioned we find a Raman scattering peak at 367 nm, for excitation energy $\lambda_{ex}$ ~ 325 nm and 417 nm, for $\lambda_{ex}$ ~ 365 nm, which is due to ZnO particles being dipolar and Raman active and can be also useful for medicine [26]

## 4. Conclusion

Bio-conjugation of metal oxide nanoparticle are important pathways to novel imaging probes and targeting agents, through our finding it is clear that FA functionalized ZnO is a better aqueous dispersed and stable system than ZnO itself as is evident from its hydrodynamic radius. The system FA-ZnO in aqueous phase shows fluorescence in the visible region that can have potential application probability for targeting biologically active tumor cells which has folate receptor proteins. It can serve to be used as fluorescent probe in biological staining and diagnostics because the substance is photochemically stable system.

## 5. Acknowledgements

The authors acknowledge the technical help received for measurements from Biophysics Division, SINP (Mr. Pulak Ray) for TEM, Surface Physics Division, SINP (Ms Tanusree Samanta) for DLS and Division of Chemical Sciences (Mr. Ajoy Das) for Fluorescence measurements. The authors also thank for the measurements of UV-Vis spectroscopy and other technical help to Soma Roy.

## 7. Tables :

Table 1: Comparison of the hydrodynamic size of synthesized pure ZnO and FA-ZnO samples Obtained from DLS

| Sample | Size (2R) estimate of pure ZnO from TEM | Hydrodynamic size (2R) of pure ZnO | Debye Layer | Size (2R) estimate of FA-ZnO from TEM | Hydrodynamic size (2R) of FA-ZnO | Debye Layer |
|---|---|---|---|---|---|---|
| Freshly prepared | 4 nm | 388 nm | 192 nm | 11 nm | 518 nm | 254 nm |
| After aging, ~ 1 month | -do- | 560 nm | 278nm | -do- | 688 nm | 388.5nm |
| After 2 months | -Do- | 560nm | -Do- | -Do- | 688nm | -do- |

Table 2: Band gap properties of ZnO from different methods

| Particle size of ZnO NPs | Method | Band gap | Reference |
|---|---|---|---|
| 41 nm | Sol-gel method, pure ZnO | 3.3 eV | 1,2 |
| 20 nm | Sol-gel method, with FA | 3.22 eV | 1,2 |
| - | Freshly prepared colloidal ZnO NPs | 3.75 eV | 2 |
| 7.5 nm | Sol-gel method, pH ~ 7 | 3.5 eV | 23 |
| 4 nm | Aqueous medium, pH ~ 7.5, pure ZnO | 3.9 eV | This work |
| 11 nm | Aqueous medium, pH ~ 7.5, FA-ZnO | 2.8 eV | -Do- |

## 8. Figures :

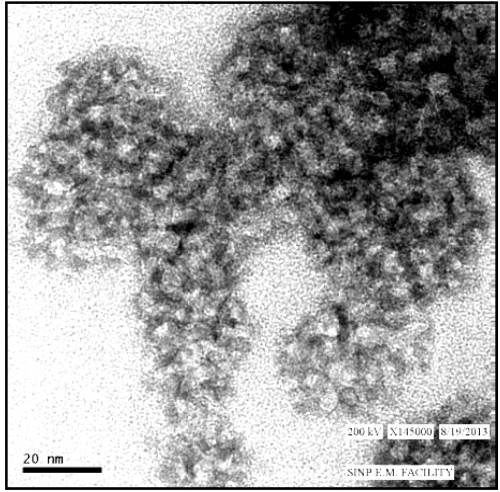

(a)

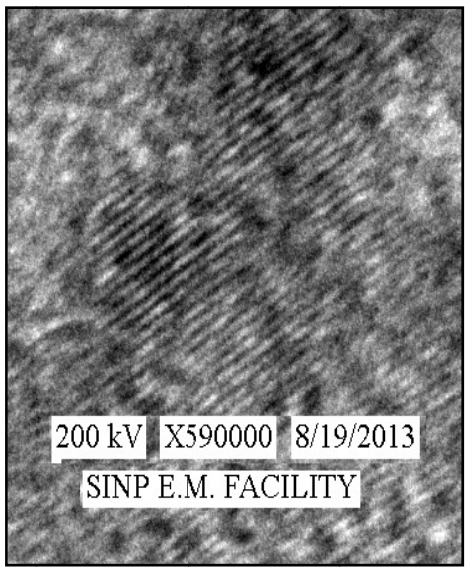

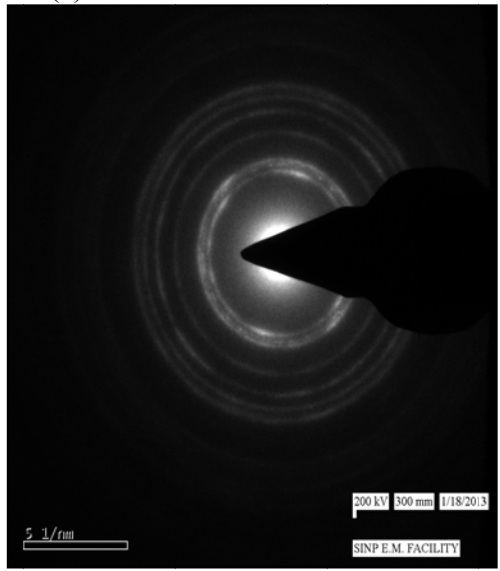

(b)

(c)
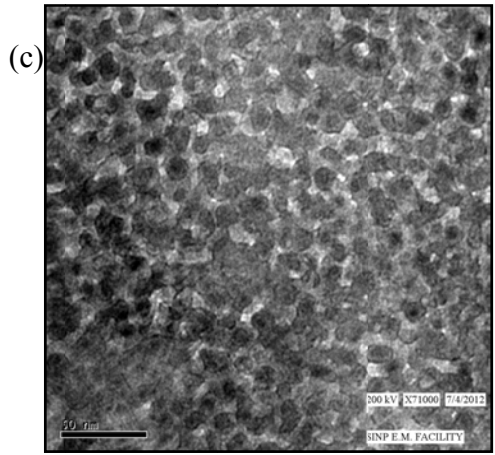

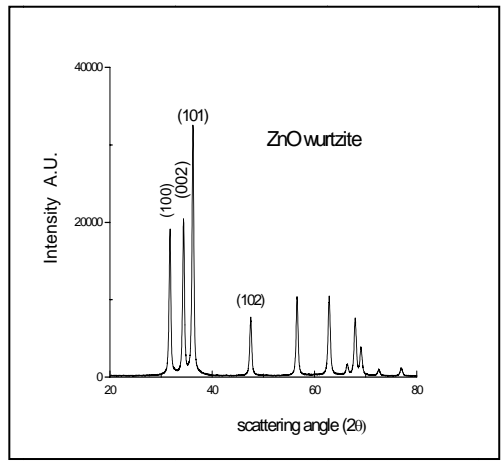

(d)

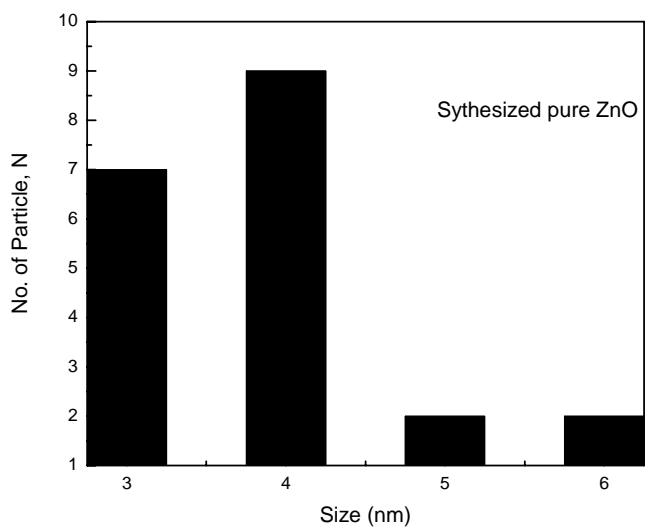

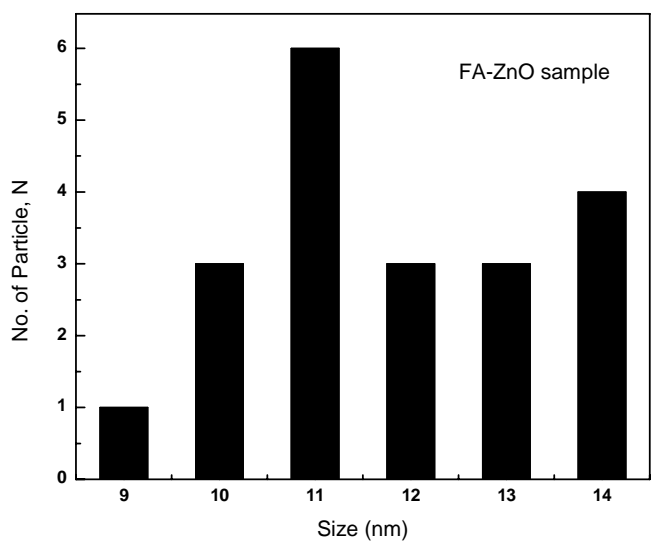

**Fig. 2.**

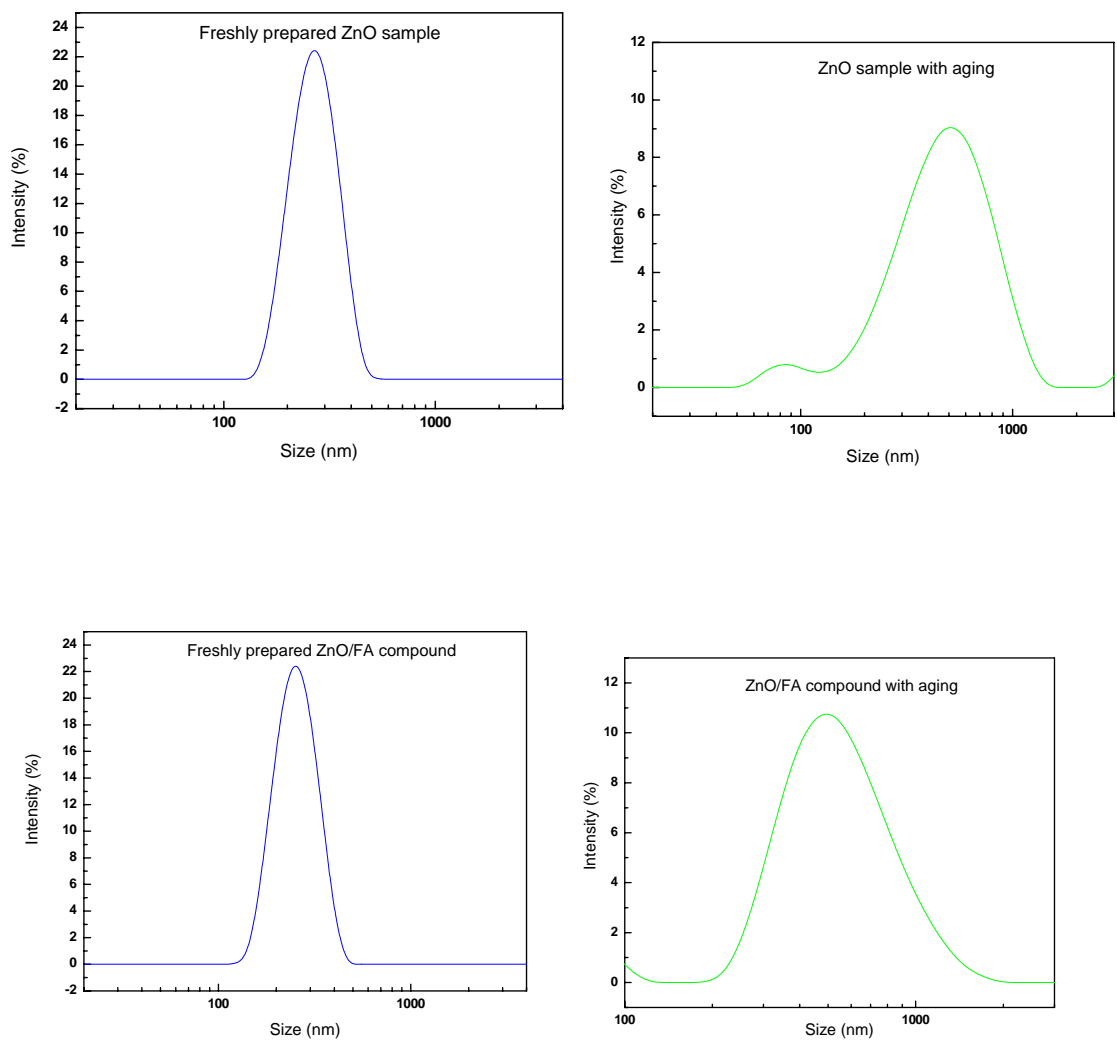

**Fig. 3.**

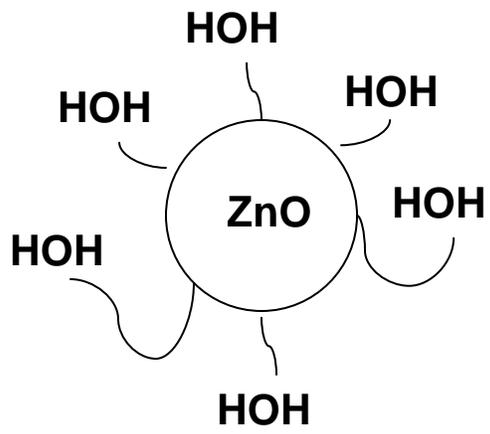 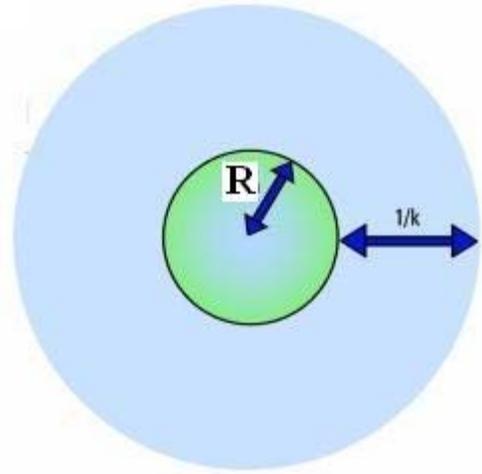

**Fig. 4.**

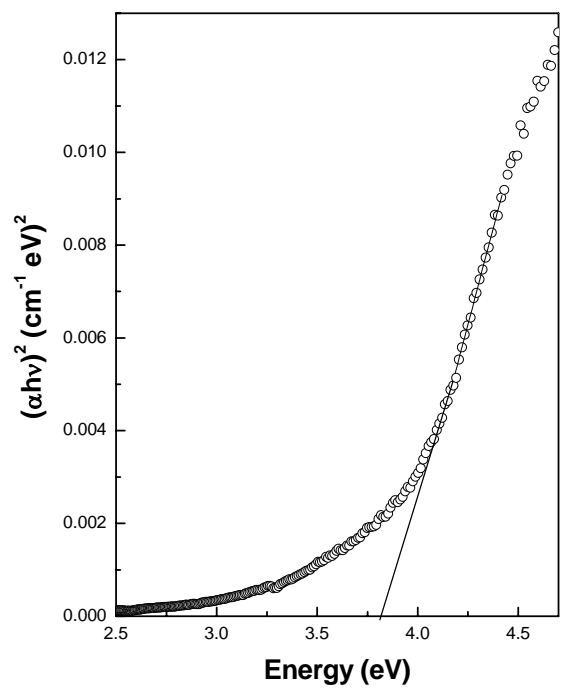

**Fig. 5.**

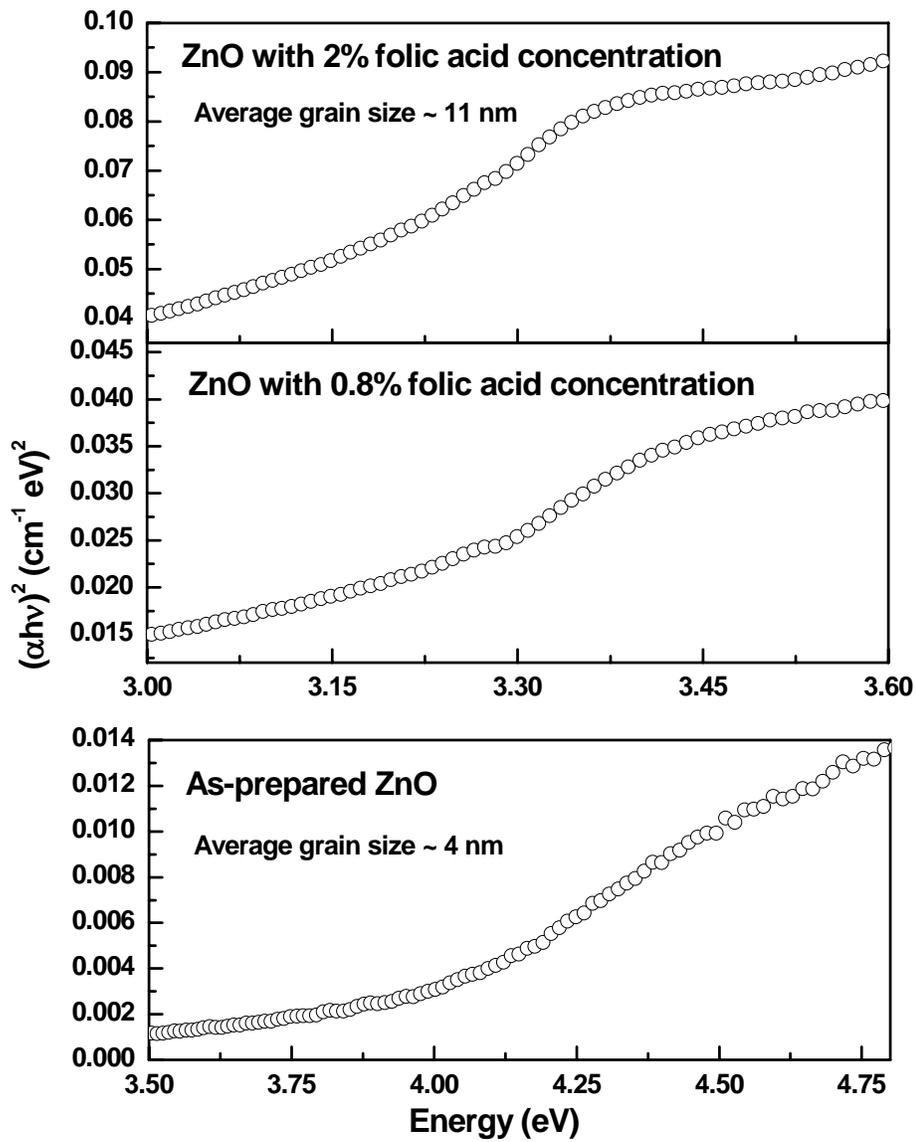

**Fig. 6.**

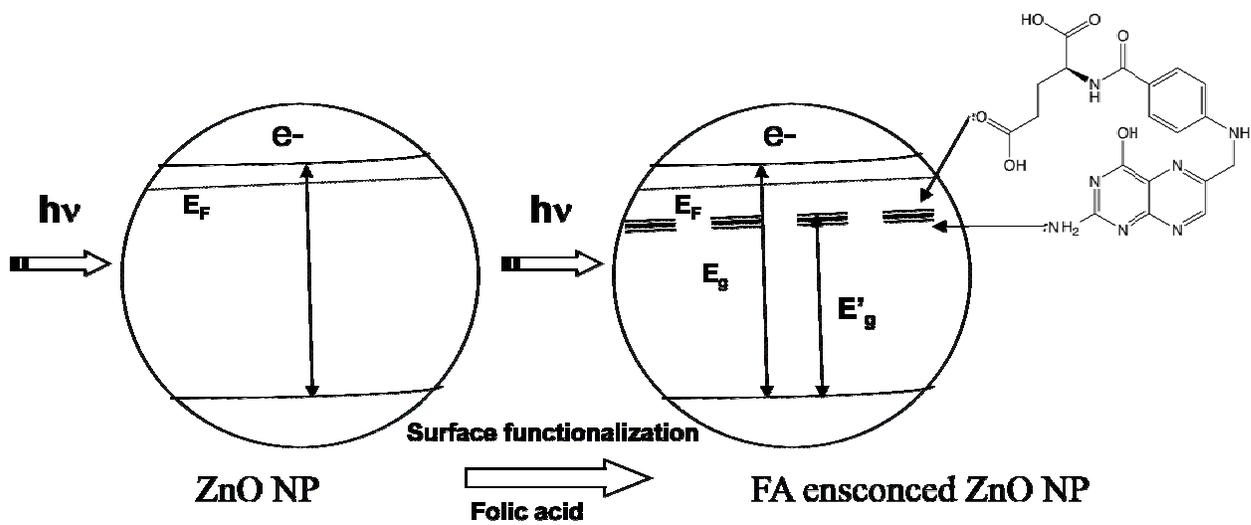

**Fig. 7.**

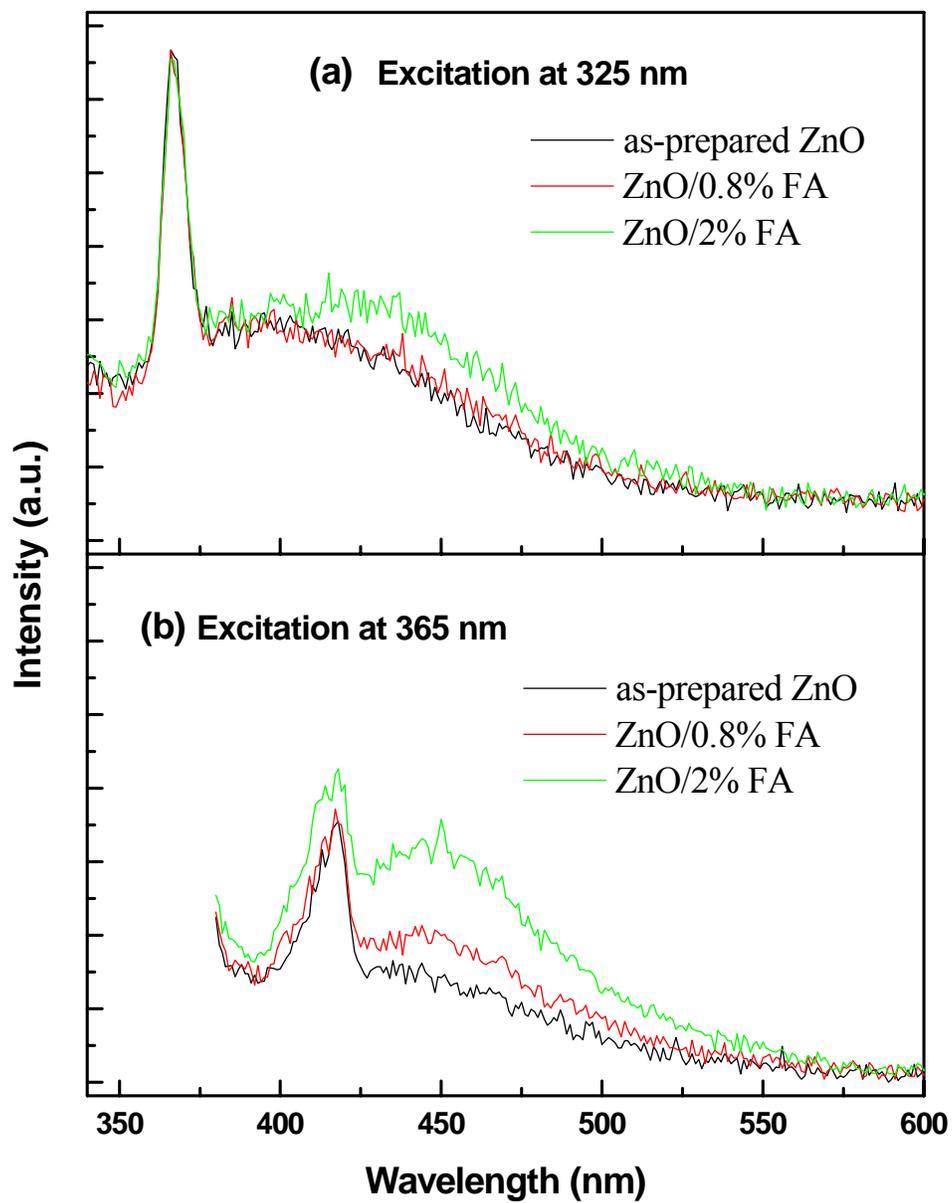

**Fig. 8.**

**Figure Caption**

Fig. 1. Transmission electron micrographs of : (a) as-prepared pure ZnO NPs , (b) the corresponding fringe structure  and also electron diffraction pattern to show the crystallinity) and (c) FA-ZnO NPs in aqueous phase.(d) X-ray diffraction pattern of ZnO.

Fig. 2. TEM  particle size distribution histograms for synthesized pure ZnO and FA-ZnO samples.

Fig. 3. Particle size distribution of pure ZnO and FA-ZnO NPs from DLS experiment.

Fig. 4. Schematic diagram of the hydrated layer of molecules (dipoles) arranged around the nanoparticle ZnO ('R'denote the radius) in TDW medium showing the Debye layer $(1/\kappa)$.

Fig. 5. A standard plot of $(\alpha h v)^2$ versus energy  for synthesized pure ZnO -NPs, showing the extrapolation to abscissa for band gap measurement.

Fig. 6. $(\alpha h v)^2$ versus energy plot of synthesized pure ZnO NPs and FA-ZnO NPs (at different concentration of FA) samples with particle size.

Fig. 7. Schematic diagram of FA ensconced charge donor level decreasing the transition level.

Fig. 8. Fluorescence spectra of synthesized pure ZnO and FA-ZnO samples at different excitation wavelength.